# VHDLBench: Benchmarking and Fine-Tuning Large Language Models for VHDL-Based Chip Design

Patrik Drazic[1], Benaoumeur Senouci[1], Boualem Benatallah[2]
University of Southern Denmark, Sønderborg, Denmark[1]
RINN AI-Research Ireland Centre for AI and Data Science, Dublin City University Dublin, Ireland[2]
senouci@sdu.dk

*Abstract*— **The rapid evolution of digital circuit complexity demands innovative approaches to hardware design automation. While Large Language Models (LLMs) have influenced software development, their application to Hardware Description Language (HDL) generation remains largely unexplored, particularly for VHDL. This paper presents a comprehensive study for AI-assisted digital circuit design, addressing gaps in current Electronic Design Automation workflows. The work introduces a VHDL dataset of 18,500+ syntax-verified examples from open-source repositories and converted Verilog designs. Five 7B-parameter LLMs were fine-tuned using parameter-efficient QLoRA technique, enabling deployment in resource-constrained environments. To evaluate model efficacy, a rigorous three-level pipeline was established: (1) syntax and synthesis validation, (2) functional simulation with testbenches, and (3) FPGA prototyping on Zynq SoCs. Results show fine-tuned models achieve up to 92% syntax correctness and 73% functional accuracy on common design tasks, with DeepSeek-Coder demonstrating the highest performance gains. Generated designs were successfully deployed on real hardware, confirming not only syntactic but operational viability. This work demonstrates that LLMs, when tailored with domain-specific data and lightweight fine-tuning can contribute meaningfully to the future of generative EDA workflows.**

## I. Introduction

### A. *Evolution of Digital Design Paradigms*

Over the past three decades, the field of digital hardware design has undergone a remarkable transformation. Digital hardware design has continually evolved through rising levels of abstraction. Early circuit design involved manual placement of logic gates and transistors, which was labor-intensive and error-prone. The introduction of Hardware Description Languages (HDLs) like VHDL and Verilog in the 1980s marked a paradigm shift, allowing designers to describe circuits at the register-transfer level (RTL) in textual form. As Moore's Law drove up circuit complexity, higher abstractions were sought to maintain productivity [1]. This led to the development of High-Level Synthesis (HLS) tools, which translate algorithmic descriptions written in languages like C or C++ into synthesizable RTL. Although C/C++ are traditionally considered low-level in software development, in the context of digital design they represent a higher abstraction compared to RTL, enabling designers to focus on algorithmic behavior rather than hardware implementation details.. HLS raised the design abstraction, but adoption has been mixed. While useful for certain algorithmic blocks, many engineers still prefer explicit RTL coding for fine-grained control over performance and resource usage. Recent studies have begun exploring use of LLMs in hardware design task [1], positing that LLMs could eventually contribute to fully automated chip design by writing correct and efficient HDL, potentially even improving design metrics like power, performance, and area (PPA). This evolution from manual RTL coding to HLS, and now to LLM-based HDL generation, represents a natural progression in abstracting and automating digital design tasks.

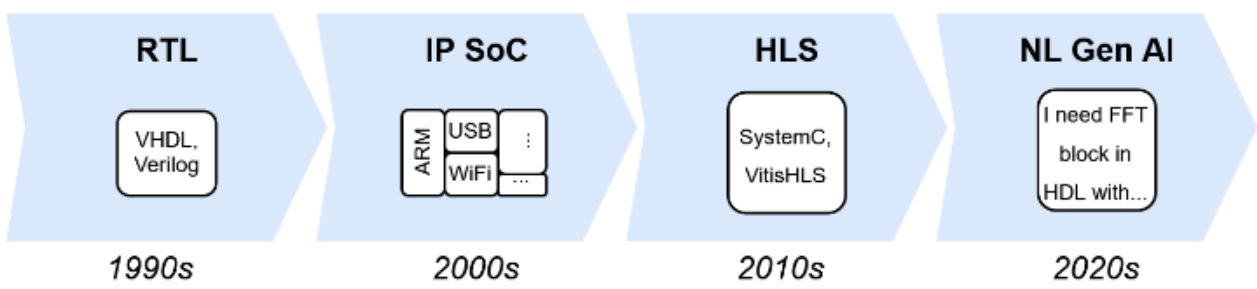

**Fig. 1. Evolution of layers of abstraction in digital hardware design: from RTL to AI-based HDL generation**

### B. *RTL and EDA in Practice*

In contemporary digital design workflows, register-transfer level (RTL) development using HDLs such as VHDL and Verilog remains central to both ASIC and FPGA design. While backend EDA tools automate synthesis and implementation, the frontend process—writing correct and optimized RTL still demands manual effort, extensive simulation, and iterative refinement. Despite advances in abstraction such as HLS, handwritten HDL continues to dominate many domains, particularly in safety-critical and low-level interface designs. This reliance on manual coding introduces productivity bottlenecks, highlighting the potential for AI-assisted design tools to enhance efficiency without compromising control.

Despite the introduction of HLS and other abstractions, many domains (especially in high-reliability and low-level interface design) still depend on handwritten RTL for maximal control. Moreover, both VHDL and Verilog communities face these challenges, with VHDL being prevalent in certain industries (e.g., aerospace, defense, European FPGA development) and Verilog in others. EDA workflows today rarely include an automated "code generator" in the loop; instead, verification feedback (e.g. compiler errors or simulation mismatches) is addressed by the engineer manually editing the HDL. This presents an opening for AI assistance, if an LLM could participate in the code-verify-debug loop, it might accelerate the workflow. In fact, researchers have started to envision such roles for LLMs in hardware design. For example, the recent LLM4EDA study categorizes potential uses of language models in EDA, including acting as an interactive design assistant, automatically generating HDL code or EDA tool scripts, and aiding in verification by analyzing or generating testbenches [1]. Overall, in practice the goal is to maintain the precision and reliability of current workflows while offloading some of the manual coding and debugging effort to intelligent automation.

### C. *AI oriented HDL Generation*

The rise of large language models (LLMs) has prompted increasing interest in their application to hardware design tasks such as HDL generation, verification, and design assistance [1], [2], [5], [6], [21], [22]. Unlike high-level synthesis (HLS) tools that translate algorithmic descriptions written in C/C++ into RTL, LLMs offer a different abstraction pathway: generating code directly in VHDL or Verilog from structured prompts or natural language descriptions. This retains RTL-level control while reducing the manual overhead typically required to write correct and synthesizable designs.

Recent studies show that LLMs can successfully generate correct HDL implementations for many fundamental tasks such as flip-flops, encoders, and simple FSMs [5], [6], [7]. However, generating more complex modules that meet functional specifications, synthesis constraints, and hardware deployability remains an open challenge. Furthermore, most existing work has focused on Verilog, leaving VHDL which is widely used HDL in aerospace, defense, and European industry relatively underexplored [8], [9].

There is also growing interest in integrating LLMs into iterative design loops with EDA tool feedback. Approaches such as AutoChip [2] and ChatEDA [21] demonstrate that LLMs can use synthesis or simulation errors to refine their outputs, potentially reducing human debugging effort. These developments point toward a future where AI agents contribute not only to HDL generation but also to iterative optimization, verification, and even co-design.

Despite these promising directions, several limitations persist in the current state of the art:

- Limited availability of large, diverse VHDL datasets suitable for supervised fine-tuning
- Absence of standardized evaluation pipelines that go beyond syntax to include simulation and hardware validation
- Variable model performance, particularly on complex design tasks

These gaps motivate the work presented in this paper, which seeks to explore the effectiveness of fine-tuned LLMs for VHDL generation using domain-specific data, parameter-efficient adaptation methods, and a multi-level evaluation strategy.

### D. *Contribution of This Work*

This study explores the capabilities of generative AI in digital circuit design, with a particular focus on direct VHDL code generation. While recent efforts such as VHDL-Eval [8] and Chain-of-Descriptions [9] have begun to explore this space, to our knowledge, this is among the first works to perform full fine-tuning of open-source LLMs specifically for VHDL generation—addressing a gap in the literature that has largely focused on Verilog.

The main contributions of this work are as follows:

- Dataset **Construction**: Diverse, verified dataset containing over 18,500 VHDL code examples
- Prompt **Design**: Prompt formatting scheme tailored for instruction-tuned models eligible for supervised fine-tuning adaptation
- Three-Level **Evaluation Pipeline**: Evaluation pipeline which includes (1) syntax and synthesizability checking, (2) functional verification using GHDL simulations and testbenches, and (3) hardware prototyping on a Zynq-based FPGA.
- Comparative Analysis: Evaluation of several fine-tuned LLMs across a diverse set of VHDL generation tasks, including basic modules (e.g., logic gates, encoders), intermediate designs (e.g., FSMs, RAM), and advanced circuits (e.g., matrix multipliers, FFTs). The study highlights architectural differences in model performance and identifies the strengths, limitations, and future improvement opportunities for LLM-based HDL generation.
- Practical Insights: Observations from model fine-tuning and evaluation, including dataset-specific challenges (e.g., VHDL syntax diversity, prompt engineering), limitations in functional verification workflows, and actionable suggestions for improving future generative HDL pipelines.

The remainder of this paper is organized as follows: Section II presents related works, highlighting the use of large language models in software and hardware code generation. Section III describes the proposed methodologies, including the dataset construction, model selection, LLM adaptation process, and evaluation strategy. Section IV discusses experimental results and benchmarking across various LLMs. Finally, Section V concludes the paper, summarizing key findings and outlining future research directions.

## II. RELATED WORK

Recent research has explored the use of large language models (LLMs) for programming assistance, code synthesis, and automated hardware design. While LLMs have shown strong performance in software programming languages, their application to hardware description languages (HDLs), , introduces new challenges due to differences in concurrency, timing semantics, and synthesis requirements. This section summarizes prior efforts in code generation for both software and hardware domains, and positions this work within that landscape

### A. *LLM for software development*

Recent advances in large language models have significantly impacted software development, enabling code generation and task automation across languages such as Python, C/C++, and Java [3]. Several instruction-tuned models and APIs (e.g., Codex, CodeLlama, DeepSeek) have been fine-tuned specifically for general-purpose programming.

However, these efforts primarily target sequential software languages and do not address the structural and timing constraints inherent in digital hardware design. The majority of existing code generation benchmarks and models ignore hardware description languages (HDLs), which demand concurrent semantics, cycle-accurate behavior, and synthesizability. As a result, applying general-purpose code LLMs to Verilog or VHDL remains nontrivial and underexplored.

### B. *LLMs for Electronics Design Automation*

With the advancements in large language models (LLMs), the application of AI in electronic design automation (EDA) has gained significant popularity, particularly for hardware description language (HDL) generation [4], [5] and digital hardware design [6]. Modern research highlights additional use cases for LLMs in EDA, such as assistant chatbots, HDL code verification, and analysis.

In the work by Thakur et al. [5], the authors fine-tuned multiple CodeGen LLM variants for Verilog HDL generation and evaluated them on a set of 17 design tasks. These included designs spanning combinational and sequential logic such as machines, muxes, RAM blocks, adders and counters. The training dataset was sourced from open repositories, such as GitHub. Building on this, Mingjue Liu et al. [7] introduced the VerilogEval framework to benchmark LLMs on Verilog generation. Their study demonstrated the potential of supervised fine-tuning (SFT) to enhance pretrained language models by leveraging synthetic design task-code pairs. However, they observed that increasing the number of fine-tuning epochs led to overfitting, reflected by improved pass@1 scores but declining pass@5 and pass@10 metrics. This indicates that while the model became better at producing a single correct output, it lost its ability to generate multiple diverse solutions. This was similarly observed in our work which reflected contents of dataset and is discussed more in Experimentation & Results section.

Subsequently, Ehsan Degan et al. [8] [9], developed the VHDLeval dataset and framework, inspired by the VerilogEval framework, to evaluate LLMs in VHDL generation. For comparison, they employed multiple techniques, such as **in-context learning (ICL)**, and conducted SFT with a relatively small dataset of 1,000 samples indicating that future-works should perform fine-tuning with more data for better results.

TABLE I. COMPARATIVE GRADING CONTRIBUTION IN CERTAIN AREAS

| Work | Dataset | HDL Focus | Evaluation Depth | Model Fine-tuning |
|---|---|---|---|---|
| VerilogEval[7] | ++ | + | ++ | + |
| VHDL-Eval[8] | ++ | + | ++ | + |
| RTLCoder [10] | +++ | + | + | +++ |
| Thakur et.al [5] | + | + | + | ++ |
| **This work** | +++ | + | +++ | ++ |

.

This work extends the findings of previous research by creating a custom design task-code dataset for SFT open to public, utilizing various open-source codebases consisting of approximately 18,500 samples. Additionally, the effectiveness of QLoRA fine-tuning method was demonstrated as an efficient approach to enhance existing LLMs for the VHDL generation task.

Five distinct LLMs, each with approximately 7 billion parameters, were selected for comparison. To better understand the impact of architecture on performance, LLMs with varying internal designs were deliberately included. Details regarding model selection are provided in the LLM Selection section.

## III. Methodology

This section outlines the methodology used to investigate the application of generative AI to digital circuit design using VHDL. The work addresses three main challenges: (1) the lack of large, diverse VHDL datasets suitable for fine-tuning, (2) the absence of detailed evaluation strategy that go beyond syntax checking, and (3) the difficulty of generating functionally correct and hardware-deployable designs across varying complexity levels. To tackle these, we curate a high-quality VHDL dataset, fine-tune several LLMs using parameter-efficient methods, and evaluate their outputs through a three-level checking pipeline. Insights gained from this evaluation loop are used to refine both the dataset and the fine-tuning setup, ultimately improving the generative performance of the models..

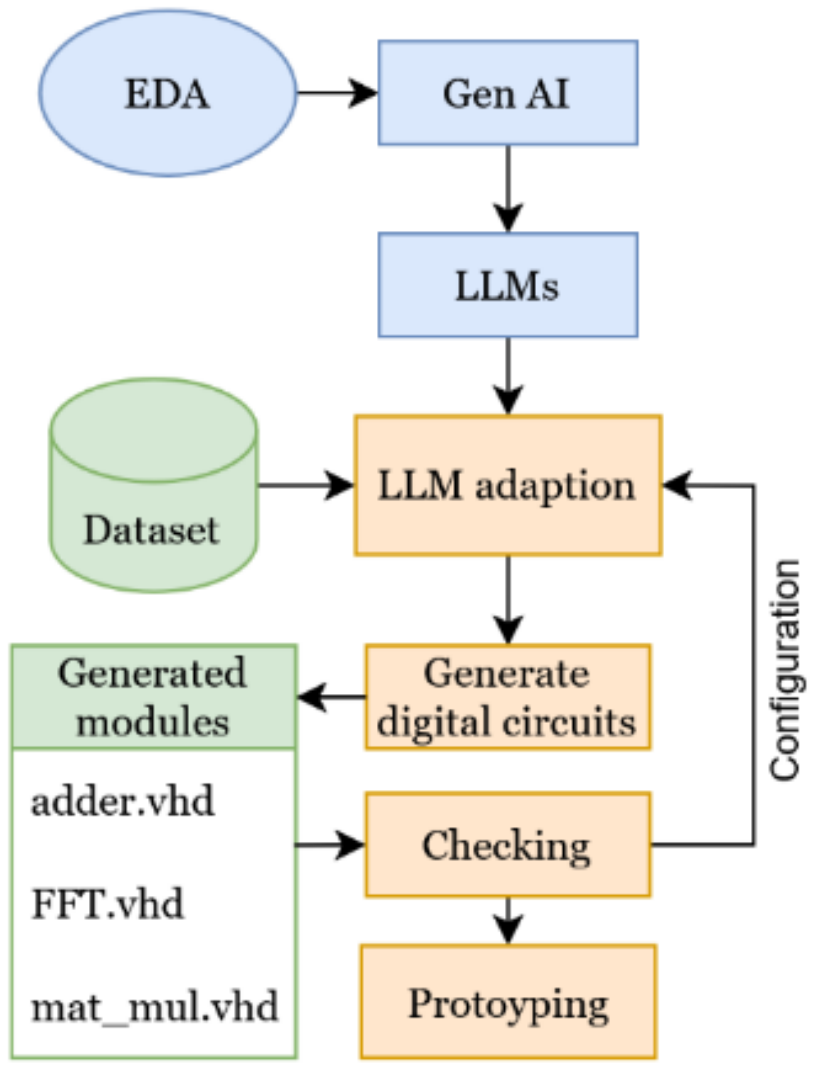

Fig. 1. Global methodology workflow

### A. *Dataset*

In this work, we present a dataset specifically curated and partially generated to enable instruction-based fine-tuning of LLMs for VHDL code generation. The dataset combines curated examples from open-source repositories and FPGA toolchains with translated VHDL examples derived from existing Verilog datasets. We also use large language models (LLMs) to generate natural language design prompts aligned with each VHDL circuit. The resulting dataset consists of over 18,500 instruction–response pairs and supports supervised training for LLMs in hardware design tasks.

#### 1) *Data Sources*

A core challenge in constructing an instruction-based dataset for VHDL is the limited public availability of high-quality VHDL examples. Unlike software programming languages, which benefit from massive open-source corpora, many hardware designs remain proprietary and closed-source. As a result, assembling a usable dataset required combining multiple sources and applying a series of preprocessing steps to ensure quality and diversity.

The primary VHDL training corpus was assembled from public GitHub repositories, following a methodology similar to that described in [5] and by translating Verilog examples from publicly available RTL-Coder [10] dataset to VHDL. This conversion was performed using a semi-automated pipeline, Verilog modules were first extracted from the RTL-Coder dataset and then translated into VHDL using the Icarus Verilog compiler. Outputs were subsequently validated using GHDL for syntax correctness, and files with structural inconsistencies or invalid syntax were filtered out.

- **GitHub repositories (via Google BigQuery)**: Custom SQL query was used to search for *.vhd* files within a snapshot of 3 million repositories. The SQL query was designed to search open gith-hub repositories and to serve as an initial filtering mechanism, incorporating criteria such as the presence of key VHDL-specific keywords (e.g., entity and architecture) and a maximum character size threshold of 10,000. This approach ensured that only substantial and relevant VHDL files were downloaded. Final corpus of VHDL examples obtained through this process comprised approximately 52,000 files.
- **FPGA development software**: In addition to this, code examples from widely used FPGA development tools were used. This includes Xilinx Vivado, Intel Quartus, and Microsemi Libero, which were locally installed. The installation directories of these tools were systematically parsed using a Python script to extract all *.vhd* files, resulting in a set comprising of approximately 4,700 unique files.
- **Hugging-Face forum dataset**: Not labeled dataset consisting of ~30,000 VHDL files was found via the Hugging Face community forums [11]. This dataset contained only VHDL code examples which again through Python script were saved as local *.vhd* files.
- **RTL-Coder dataset (converted from Verilog)**: To further enhance the dataset size while trying to have as good diversity as possible, the publicly available dataset was leveraged and project files from study [10], whose authors made their project resources accessible on GitHub. After downloading the RTL-coder Verilog dataset it was converted to *.csv* format and all rows of column containing code examples have been translated to VHDL using Icarus Verilog compiler [33]. Files that failed conversion were excluded; this provided from original 27k examples to approximately 14.400 examples which had to be further checked and filtered. This expansion of examples showed to be less impactful and output of LLMs started getting biased towards specific form of converted examples which would lead to less optimized digital circuits, this is explained further in pre-processing section.

*2) **Dataset pre-processing***

Since the collected *.vhd* files originated from diverse sources and were acquired in bulk, no assumptions could be made about the quality, completeness, or synthesizability of individual VHDL examples. To ensure the integrity and utility of the dataset for fine-tuning, several processing steps were applied. The first and most critical step was syntax validation, performed using the open-source GHDL analyzer and simulator [34]. This tool was used to parse each VHDL file to detect invalid files. The main reasons for file exclusion during syntax checking were: (1) the design was not self-contained and relied on external modules or hierarchical definitions that were missing, and (2) the presence of general syntax errors, such as incomplete statements or undeclared signals, which prevented successful compilation check by GHDL.

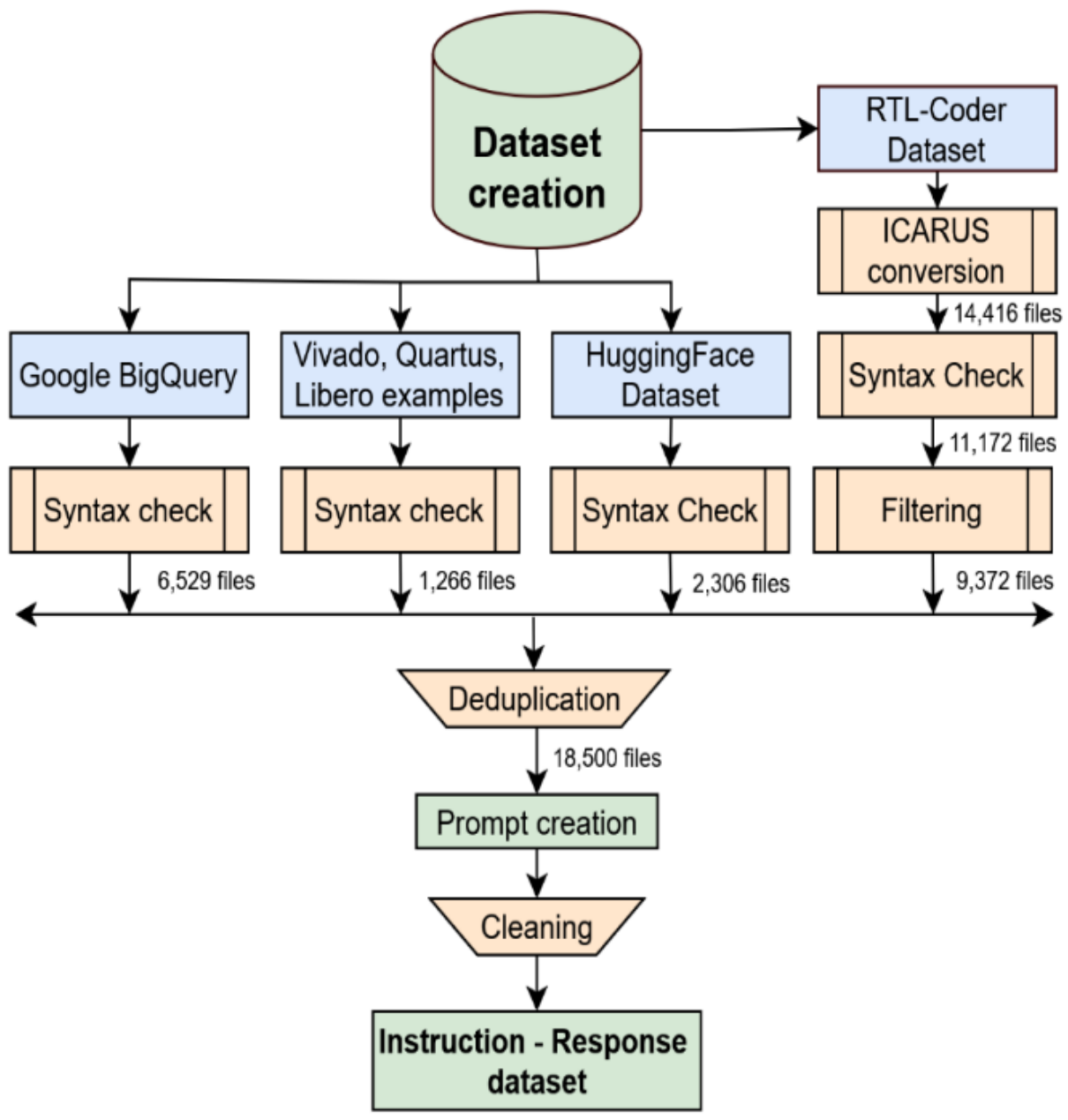

**Fig. 2. Flowchart of the dataset curation for supervised fine-tuning on VHDL generation application**

One important structural feature of the final dataset is that each example represents a self-contained VHDL circuit meaning a complete design with no dependencies on external modules or hierarchical composition. This includes examples such as a 4-bit comparator, an 8-bit shift register, or a simple traffic light controller, all encapsulated within a single entity-architecture pair. While this structure simplifies evaluation and fine-tuning, it also introduces a limitation: generating hierarchical designs, such as systems composed of multiple interacting submodules under a top-level design, remains an open challenge for current LLMs and a direction for future work.

As a result, the dataset was significantly reduced in size:

- GitHub (BigQuery): reduced from ~52,000 to ~6,500 valid files
- Software examples: reduced from ~4,700 to ~1,266
- Hugging Face dataset: reduced from ~30,000 to ~2,300
- RTL-Coder conversions: The RTL-Coder dataset initially contained approximately 26,000 Verilog examples, which were converted to VHDL using Icarus Verilog. This conversion process reduced the total to 14,416 VHDL files. After performing syntax checks using GHDL, the number was further reduced to 11,172 valid examples. Additional filtering was performed, as Icarus Verilog occasionally produced suboptimal or incorrect VHDL code. In particular, some files included excessive use of intermediate signals and temporary variables—most notably, repeated instances of key words such as *signal tmp_ivl*. These examples often deviated structurally from standard VHDL practices, making them difficult for engineers to interpret and inconsistent from a learning perspective for LLMs. To filter out such cases, files containing more than nine occurrences of the keyword tmp_ivl were excluded. After this filtering step, the final curated set from the RTL-Coder conversions consisted of approximately 8,300 VHDL examples..

Additionally, simulation and testbench files were manually removed to focus on synthesizable VHDL designs. Following this, an approach inspired by [5] was employed, utilizing MinHash and Jaccard *similarity* algorithms to perform a 95% deduplication of the dataset and remove near identical files to prevent biases. This process ensured a high level of diversity among the VHDL examples and a broad range of design patterns and complexity levels.

As an additional (optional) step, incorporating the VHDL Style Guide (VSG) tool can be highly beneficial. VSG is available both as a command-line tool and a Python package that automatically reformats a given *.vhd* file. It analyzes the file for style guide violations such as incorrect or missing indentation, inconsistent spacing, and deviations from naming conventions and corrects these issues. By enforcing a standardized coding style, VSG ensures that the generated VHDL code is both consistent and readable, thereby enhancing maintainability and facilitating easier debugging. This was used only on examples acquired from *Huggingface* dataset.

### 3) *Prompt Processing*

To enable supervised instruction tuning of LLMs for VHDL code generation, each circuit in the dataset had to be paired with a corresponding natural language (NL) description referred to as prompt. These prompts serve as task instructions, i.e., simulated user requests for a specific hardware module.

Because the majority of collected VHDL files were not originally associated with textual descriptions, a prompt generation process was applied. Specifically, each VHDL example was passed to a commercial LLM (initially GPT-4o-mini, later Gemini 2.0 Flash), configured to summarize the circuit behavior as a short NL design task. This was implemented via API calls using a structured system prompt, including few-shot examples for in-context learning (ICL) [8 [12], to ensure output consistency and alignment with the desired prompt structure.

The generated prompts were required to meet the following criteria:

- Include the entity's input and output port names and types,
- Describe key behavioral aspects (e.g., output conditions or control logic),
- Avoid code-specific terminology or implementation details.

An example of a generated instruction(prompt) is shown in *Dataset Statistics* subsection. In Figure 3 proess of using API for creating prompts is shown. Where circuit design in VHDL is given to advanced LLM to summarize the given design in sentences as a design task problem description,

After prompt generation, code comments were removed from the VHDL files to reduce sequence length and token overhead during model training. This step ensures compatibility with LLM context window limits while retaining the syntactic and semantic integrity of the code. Figure 2 illustrates the complete prompt creation process.

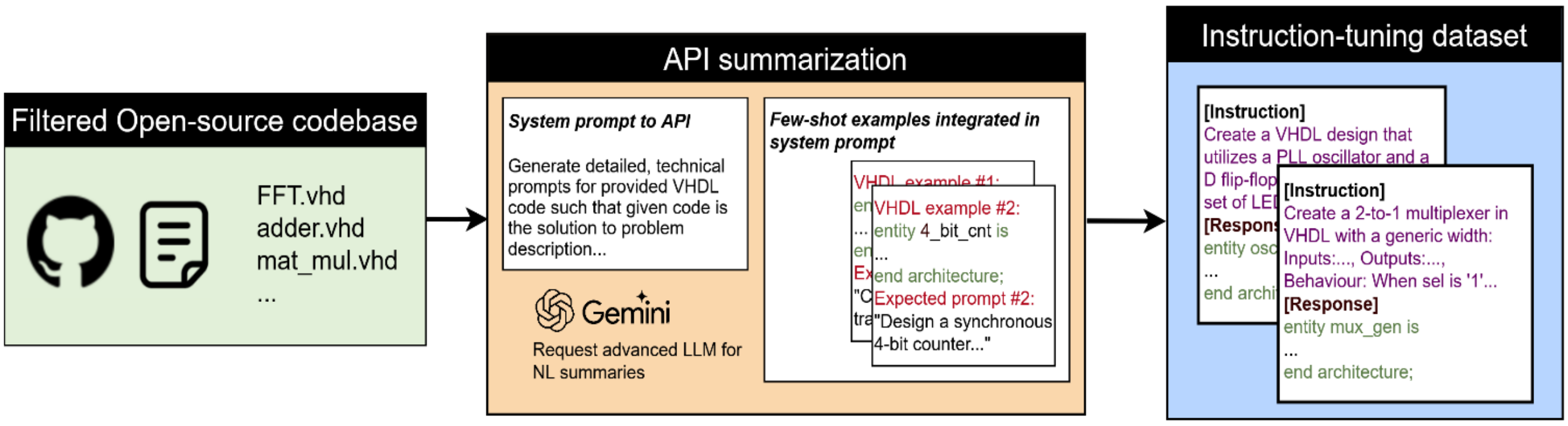

**Fig. 3. Creation of prompts by giving VHDL examples to advanced LLM**

### 4) *Dataset Statistics*

Given the large number of the collected VHDL examples, it was essential to gain a clearer understanding of the dataset's overall structure and complexity. To clarify the classification process, it's helpful to provide examples of what was considered basic, intermediate, or advanced complexity. Specifically, designs with fewer than 5 ports in the entity were classified as basic, those with 5 to 9 ports as intermediate, and entities with more than 9 ports as advanced. The classification also took into account the types of used signals and ports.

To estimate the complexity of each VHDL design, the Gemini API was used to perform an automated classification. For each example, the entity declaration was extracted and passed to the API, which then determined the design's complexity based on the number, type, and structure of ports. Basic designs typically included simple circuits such as D flip-flops or multiplexers with just a few scalar signals (e.g., clock, reset, single-bit data). Intermediate designs involved modules like 8-bit register files or small ALUs, featuring a mix of scalar and vector signals. Advanced designs were more complex, with over 9 ports including multiple data buses and control signals, as seen in modules like UART controllers or AXI interfaces.
This method is not accurate but has provided rough estimate of types of examples in dataset, which can be see in result section after evaluating fine-tuned LLMs.

As shown in Figure 4, the majority of the dataset consists of basic (56.2%) and intermediate (31.9%) circuits, with advanced designs making up 11.9%. This imbalance reflects the reality of publicly available VHDL code, where simple modules are more common.

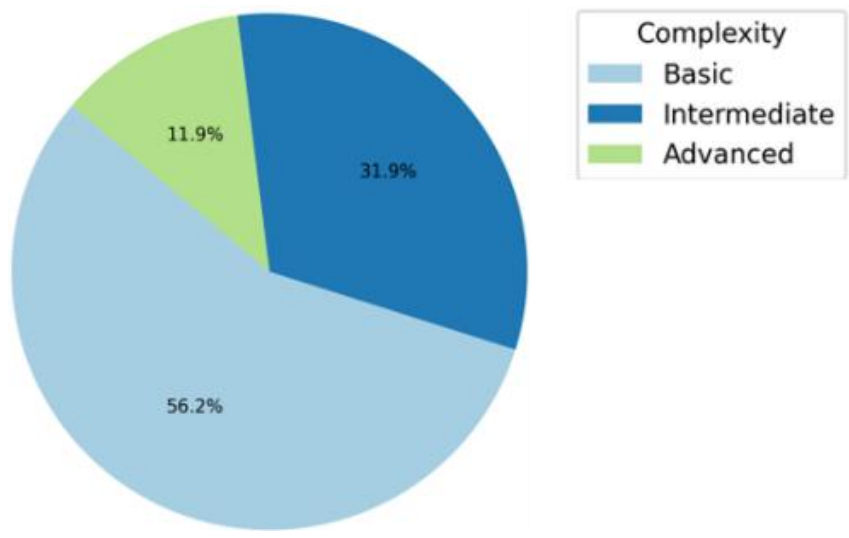

**Fig. 4. Distribution of dataset examples by circuit complexity class**

In addition to complexity labeling, each VHDL code samples were tokenized using the GPT-2 tokenizer to determine its token length. Useful metric when considering LLM context window constraints during fine-tuning.

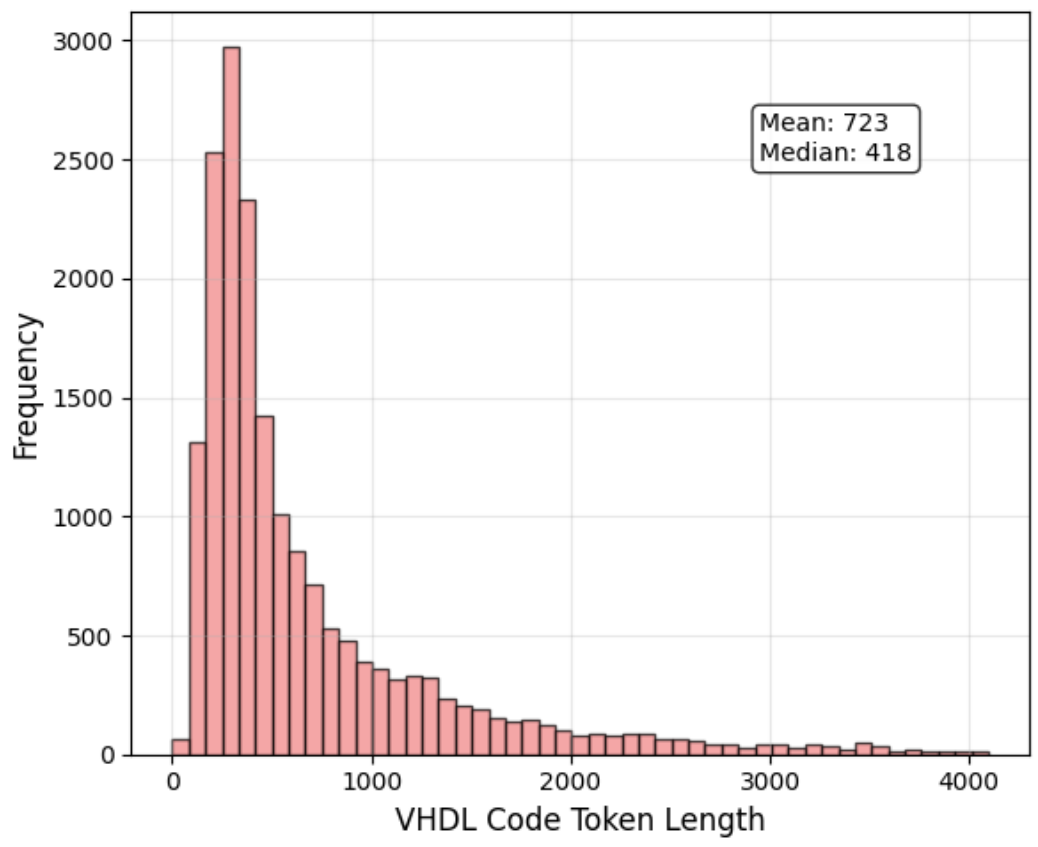

**Fig. 5. Histogram of VHDL code token lengths (GPT-2 tokenizer)**

As shown in Figure 5, most examples fall within the 200–1000 token range, with a mean of 723 tokens and a median of 418. Although a small number of examples exceed 2000 tokens, the vast majority of designs remain well within the maximum 4096-token limit used during training.

*Example of dataset entry pair*

```
[ INSTRUCTION ] :
Create a 5-input OR gate in VHDL with:
Inputs: I0, I1, I2, I3, I4 (all std_ulogic)
Output: O (std_ulogic)
Behavior: The output O should be high if at least one of the inputs
I1, I2, I3, or I4 is high; however, if input I0 is low and any of the
other inputs are high, O should also be high. Ensure to utilize
proper logical operations for this behavior.
[ RESPONSE ] :
library IEEE;
use IEEE.STD_LOGIC_1164.all;
entity OR5B1 is
  port(
    O : out std_ulogic;
    I0 : in std_ulogic;
    I1 : in std_ulogic;
    I2 : in std_ulogic;
    I3 : in std_ulogic;
    I4 : in std_ulogic
    );
end OR5B1;
architecture OR5B1_V of OR5B1 is
begin
  O <= ((not I0) or I1 or I2 or I3 or I4);
end OR5B1_V;
```

As shown in the previous example, each dataset entry is structured as an instruction–response pair, where the instruction is a natural language description of the desired circuit, and the response is a complete VHDL implementation.

## *B. Large Languange Model Selection*

To address the task of generating VHDL code from natural language prompts, this work employs several large language models (LLMs) that are pre-trained on source code and instruction-tuned for better alignment with human inputs. Choosing appropriate models is a critical step, as it directly impacts the quality, usability, and reproducibility of the overall workflow.
The selection of models was guided by the following criteria:

- Code-oriented models were preferred over general-purpose chatbots to ensure better familiarity with programming syntax and logic structure
- Instruction-tuned versions of these models were chosen to match the instruction–response format of the dataset
- Open-source availability was prioritized to ensure reproducibility and reduce the dependency on commercial APIs
- Model size needs to be weighted with hardware availability

TABLE II. BASELINE LLMS USED IN THIS STUDY

| Used Model | Model Architecture | Context window |
|---|---|---|
| Mistral-7B-Instruct-v0.3 | MistralForCausalLM | 32k |
| falcon-7b-instruct | FalconForCausalLM | 2048 |
| granite-8b-code-instruct-4k | LlamaForCausalLM | 4096 |
| CodeLlama-7b-Instruct-hf | LlamaForCausalLM | 16k |
| deepseek-coder-6.7b-instruct | LlamaForCausalLM | 16k |

## *C. LLM Fine-Tuning*

To adapt general-purpose language models for the task of VHDL code generation, each selected LLM was fine-tuned on the curated VHDL dataset described earlier. Given the size of the models and the hardware limitations typical of academic or local environments, a parameter-efficient fine-tuning approach was employed.

This work utilizes QLoRA (Quantized Low-Rank Adaptation), a method that enables large-scale model fine-tuning with significantly reduced memory footprint. Rather than updating all model parameters, QLoRA freezes the base model and trains only a small number of LoRA (Low-Rank Adapter) layers inserted into the transformer architecture. This adapter-based strategy drastically lowers the computational requirements while still achieving substantial performance gains comparable to full parameter fine-tuning.

In addition, 4-bit quantization was applied to compress the model weights during training, further reducing memory overhead. This made it feasible to fine-tune 7B-parameter models using a NVIDIA A100 GPU (40 GB VRAM). Training was conducted using the Hugging Face ecosystem. Each model was fine-tuned for 5 epochs with average training time of 6 hours.

Moreover, several experiments were conducted to evaluate the effect of training strategies on model performance. Studies such as [13] and [14] have investigated the effects of including prompt token loss versus masking those tokens during training. Their results indicate that for long prompts and short answers, masking prompt tokens yields better performance. However, in our testing, the performance remained largely the same. While literature suggested masking improves performance for long prompts and short answers, in this task the difference was minimal. Only interesting observation was that training on the full input sequence (including the prompt) occasionally helped the model learn contextual relevance more effectively. This was particularly useful for vague or underspecified prompts, where the model, having seen more detailed prompts during training, would sometimes generate missing context or elaborate on the instruction before producing the corresponding VHDL code. Additionally, the NEFTune parameter [15] which ads noise to the embedding vectors during fine-tuning was tested to evaluate its effect on the fine-tuned Mistral LLM. Setting it to 5 resulted in a 3% degradation in accuracy. However, further human evaluation would be necessary, as the noise parameter may contribute to improved robustness when handling diverse prompt types during inference.

After exploring hyper-parameters and different techniques, it became evident that the most significant if not the only relevant factor for improving LLM generalization in this study was the dataset itself. Consequently, the fine-tuning method and parameters were kept mostly consistent across all models.

### 1) *EvaluationEvaluation Design-Task Set*

To evaluate LLMs on the task of VHDL generation from natural language (NL), a separate set of design tasks was constructed. Since the VHDL-Eval dataset from [8] was not publicly available, we created our own evaluation sets based on examples from *HDLBits.net* and *FPGA4Student.com*. Prompts were generated using the same approach as in the training dataset, but additionally paraphrased to ensure the uniqueness of user inputs to the LLM. These refined prompts (NL design tasks) were then used as inputs for the LLMs during inference.

Two evaluation sets were used in this study:

- **100-eval set**: A diverse collection of 100 design tasks of varying complexity, intended to provide a broad performance overview of the LLMs
- **15-eval set**: A smaller, subset consisting of 15 tasks (5 basic, 5 intermediate, and 5 advanced designs). This subset was used for in-depth analysis, including functional simulation and hardware prototyping. The reduced size was necessary due to the labor-intensive task of testing the functionality of a large number of generated circuits and performing hardware prototyping

In Figure 7, process of creating evaluation set is shown.

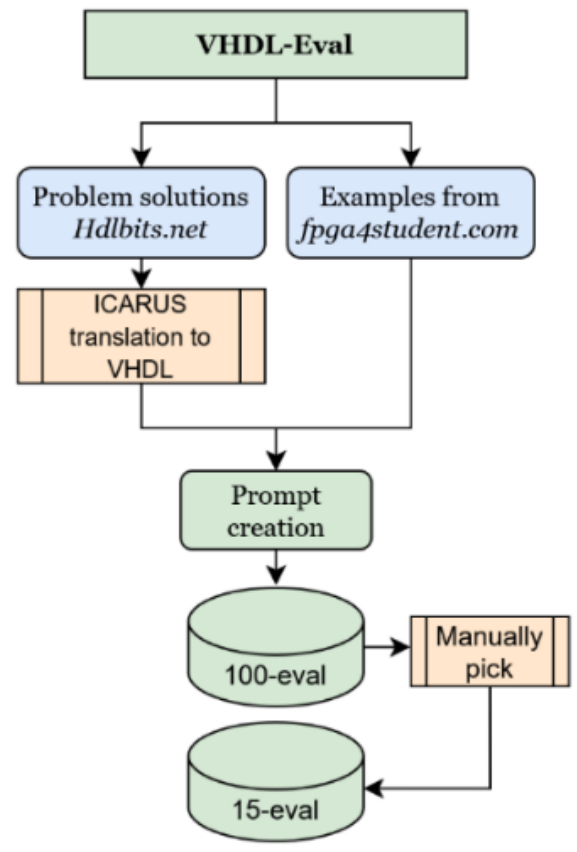

**Fig. 6. Eval-dataset creation flow-chart**

And example of paraphrased prompt used during inference:

*Example problem description paraphrased, used during inference*

[ INSTRUCTION ] :
*"Develop a VHDL module for a 9-to-1 multiplexer with a 9-bit std_logic_vector input called d and a 4-bit selection signal, sel, while producing a single std_logic output named out_sig. The design should select one bit from d based on the binary value of sel. For example, if sel equals "0000", then out_sig should be assigned d(0); if sel is "0001", then out_sig should equal d(1); this pattern continues until sel is "1000", which selects d(8). For any other value of sel, the output should default to '0'."*

### 2) *Three-Level Validation Pipeline*

When evaluating LLMs for digital circuit design, the assessment of generated output can be categorized into three key stages, as described in [1]:

1. **Syntax correctness** - The generated RTL code must have to proper syntax and be verified to ensure it can be synthesized into a netlist.
2. **Functionality accuracy** - The circuit should perform according to the user's expectations. This is validated using test-benches that cover all cases of possible inputs.
3. **Hardware prototyping** - while not real PPA assessment, in this work the quality of the circuit itself was investigated during hardware prototyping (used LUT's, power usage, timing constraints from Vivado)

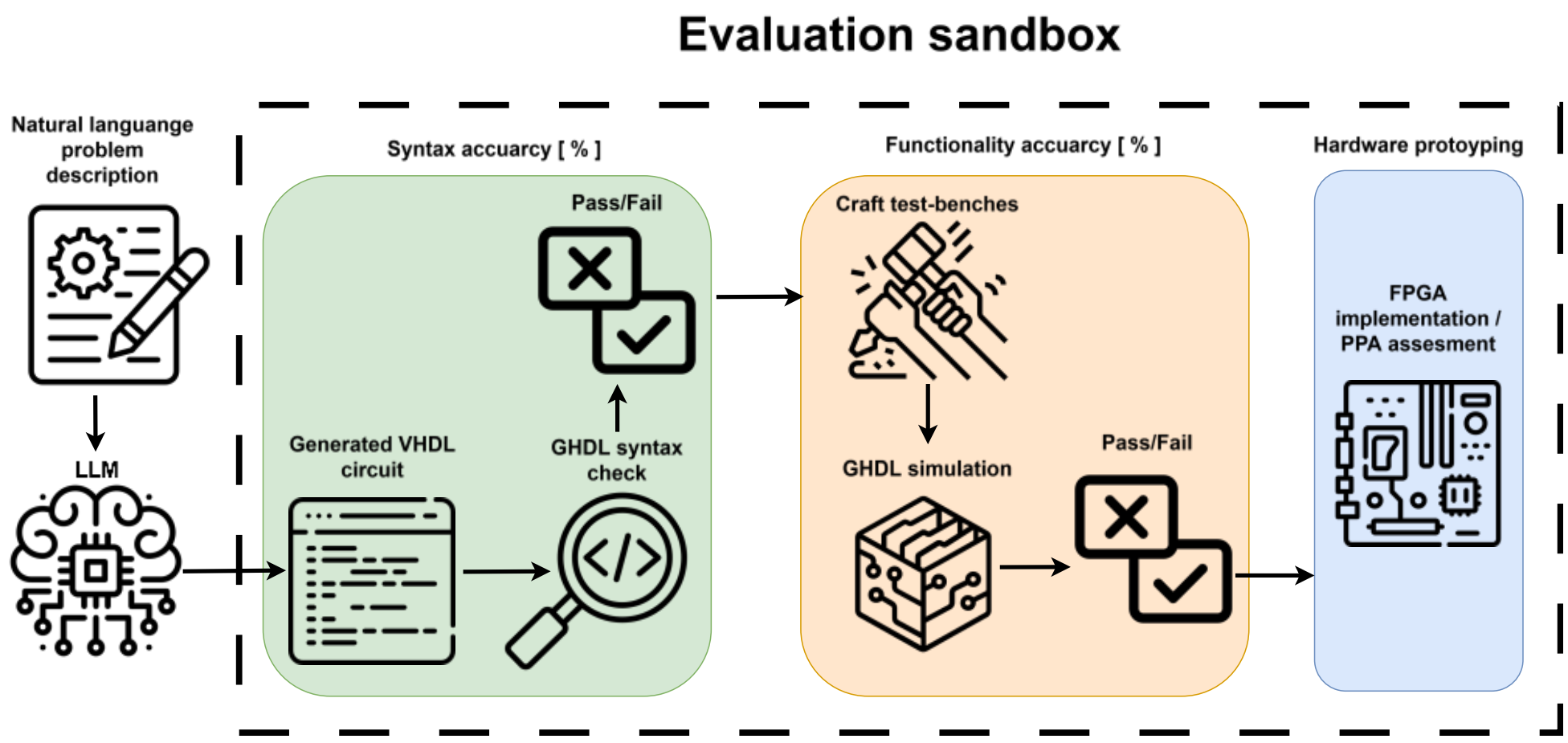

**Fig. 7. High-level custom workflow for assesing and evaluating LLMs**

This multi-level evaluation process was essential, as relying solely on syntax or synthesis checks is insufficient for assessing the practical utility of LLM-generated circuits. While many generated circuits may pass initial syntax validation and can be synthesized , they can still contain flawed or unintended logic. To detect such errors, functional simulations were conducted using testbenches that stressed each design under a range of input conditions. Moreover, hardware prototyping provided further insight into the deployability and efficiency of the generated circuits

**Syntax Checking** was performed using an automated pipeline. During inference, every LLM-generated response was truncated at keywords such as *end;* or *end architecture;* and saved as individual *.vhd* files. After the LLM completed inference on all evaluation prompts (generating *n* completions per prompt), these files were collected in a designated folder. A Python script then systematically validated each file using GHDL's syntax checking functionality with the command *ghdl -s filename.vhd.* The evaluation metric was binary: files that compiled without errors were marked as syntactically correct, while those producing compilation errors were marked as failed. This automated process generated logs of successful and failed compilations, enabling calculation of syntax correctness rates. Only files that passed this automated validation stage were advanced to functional verification.

**Functional Verification** was performed manually and applied only to the 15-problem evaluation set due to its labor-intensive nature. Files that passed syntax validation were first processed through a deduplication algorithm to identify unique VHDL implementations for each prompt. Since LLMs often generated identical solutions for the same prompt across multiple completions, this step ensured that only different designs were evaluated. For each unique VHDL solution, testbenches were created manually using ChatGPT to generate the initial testbench skeleton, followed by manual insertion of assert statements tailored to verify specific circuit behaviors. Each evaluation step consisted of three components in a folder: the LLM-generated VHDL circuit, the corresponding testbench, and a *Makefile* for automated simulation. The simulation process used GHDL commands (*ghdl -a, ghdl -e, ghdl -r*) with a fixed simulation stop time of 100µs. Functional correctness was determined through manual evaluation of terminal-based simulation outputs, analyzing both assert statement results and circuit behavior against expected functionality. This manual verification process constituted approximately 90% of the total evaluation time and results were systematically recorded in evaluation tables. A design was considered functionally correct only if it consistently demonstrated the intended behavior throughout all test scenarios without assertion violations.

*Example of generated circuit simulation (singlePort RAM), and outputs used to validate circuit functionality*

```
[ SIMULATION OUTPUT ]:
ubuntu@desktop:~/_TESTING_/singleportRAM$ make
ghdl -a generated_ram.vhd
ghdl -a -fsynopsys generated_ram_tb.vhd
ghdl -e tb_RAM_128x8
ghdl -r tb_RAM_128x8 --stop-time=100us
generated_ram_tb.vhd:70:8:@0ms:(report note): Writing 0xAB to address 0x01
generated_ram_tb.vhd:76:8:@10ns:(report note): Writing 0xCD to address 0x02
generated_ram_tb.vhd:82:8:@20ns:(report note): Reading from address 0x01, expecting 0xAB
generated_ram_tb.vhd:85:8:@30ns:(report note): Actual value read: 10101011
generated_ram_tb.vhd:92:8:@30ns:(report note): Reading from address 0x02, expecting 0xCD
generated_ram_tb.vhd:95:8:@40ns:(report note): Actual value read: 11001101
```

**Hardware Prototyping**. To examine hardware feasibility, functionally correct designs were synthesized and implemented using Xilinx Vivado on a Zynq SoC platform [32]., This additional test confirmed the direct portability of the generated VHDL to physical hardware. Post-implementation reports from Vivado were used to extract metrics relevant to the physical design, including:

1. **Power** – The power consumption was estimated based on the dynamic power value provided by Vivado, which accounts for the power consumed by signals, logic elements, and I/O activities within the circuit
2. **Performance** - The performance was evaluated through timing analysis. The total propagation delay was determined from the longest combinational path identified in the default timing report generated by Vivado.
3. **Resources** – The resource utilization was reported in terms of the total number of used configurable logic blocks (CLBs), including look-up tables (LUTs), flip-flops (FFs), DSP slices, RAM blocks and I/O's

### *3) Checking analyis and used metrics*

Well known and used *pass@k* metric [16] was used. The *pass@k* metric, a standard in related research, quantifies the

probability of obtaining a correct output within the top-k attempts. The metric is formally defined as:

$$pass@k = \mathbb{E}_{problems}\left[1 - \frac{\binom{n-c}{k}}{\binom{n}{k}}\right] \qquad (1)$$

Where in this case $n \geq k$ represents the total number of generated outputs per prompt, and c is the number of correct outputs for the same prompt. This definition accounts for scenarios where a correct solution in this case, syntax and/or synthesis correctness may exist within the top-k generated outputs, but not necessarily at the first position.

*Algorithm 1: pass@k*

```
Input: num_samples (n), num_correct (c), k
Output: pass@k value
import numpy as np
    def pass_at_k(n, c, k):
        if n - c < k : return 1.0
        return 1.0 - np.prod(1.0 - k / np.arange(n - c + 1, n + 1))
```

## IV. Inference setup & Results

### A. *Inference setup*

During inference, models' generation configuration was carefully controlled to ensure reproducible results. Each LLM input, being design task description in natural language had fixed sampling temperature (t) of $t = 0.1$, as this value has been shown to yield more deterministic outputs, which is essential for code generation tasks [7]. Higher temperature values increase the risk of generating syntactically incorrect HDL code due to increased randomness in token selection. The number of completions per design task was set to $n = 3$ for the 100-eval set, serving as the minimum required to assess accuracy using *pass@k* metrics. For the 15-eval set, $n = 10$ completions were generated to provide a more accurate estimation of the probability of obtaining a correct output.

### B. *Perfomance Analysis and Results*

Figures 8 and 9 depict graphical representations of the average pass@1 metric for syntax and synthesis accuracy, respectively when evaluating LLMs with 100-eval set. The pass@1 was calculated for every generated VHDL designs from given design task description. The average pass@1 value was then taken to illustrate general overview of each model's performance.

Looking at the baseline performance before fine-tuning, significant variations were observed across different model architectures. CodeLlama-7b-instruct achieved the highest baseline performance at 68%, which can be attributed to its specialized pre-training on code repositories including various programming languages. In contrast, Falcon showed the poorest baseline performance at merely 3%, indicating that its general-purpose architecture was poorly suited for the specialized task of HDL generation.

The impact of fine-tuning varied significantly across models, revealing interesting patterns about model architecture and adaptability. DeepSeek-Coder demonstrated the most dramatic improvement, jumping from 57% to 92% syntax correctness. CodeLlama's improvement from 68% to 81% was more modest, potentially indicating some degree of saturation given its already high baseline performance.

Another indicator of successful fine-tuning was observed during a human evaluation phase. A subset of VHDL outputs that failed both before and after fine-tuning were examined by electronics engineering student (the author), using GHDL error messages to identify syntax violations. The differences between pre- and post-finetuned outputs were substantial. In many cases, only minimal edits were needed to correct syntax for outputs after fine-tuning, demonstrating a significant improvement in the model's ability to produce near-valid VHDL code, even when exact syntax correctness was not achieved.

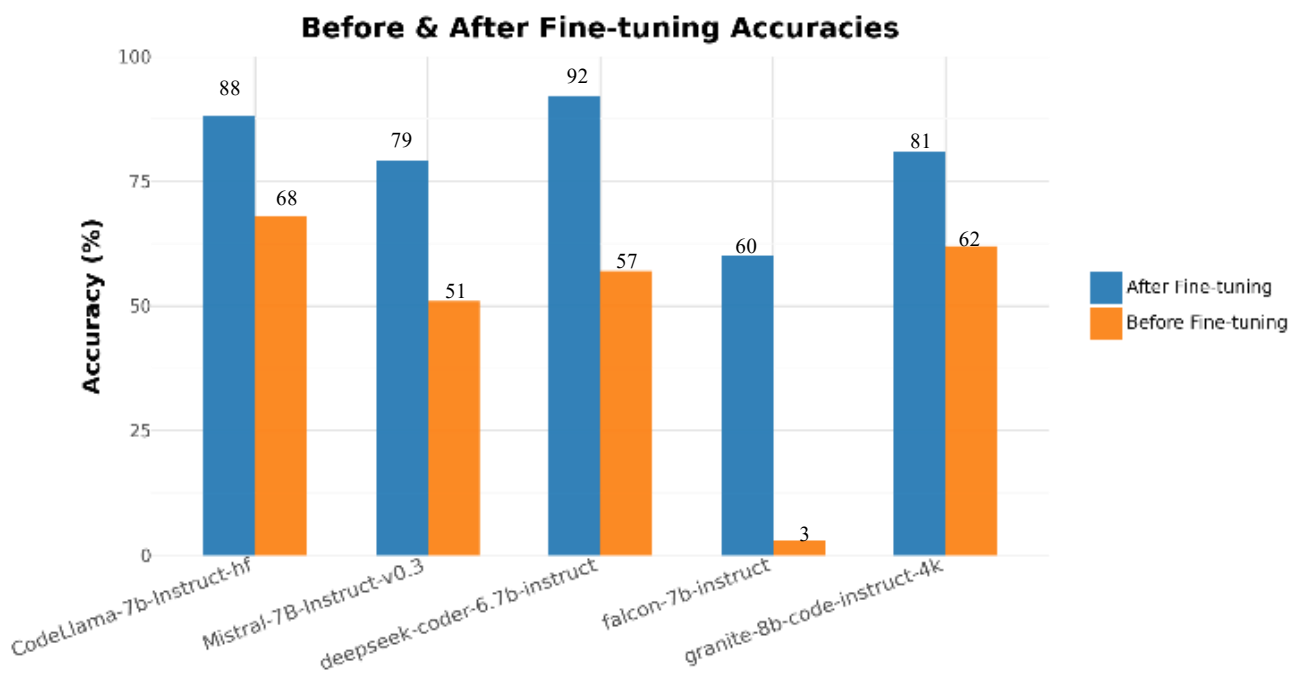

**Fig. 8.** General accuarcy in generating designs with correct syntax

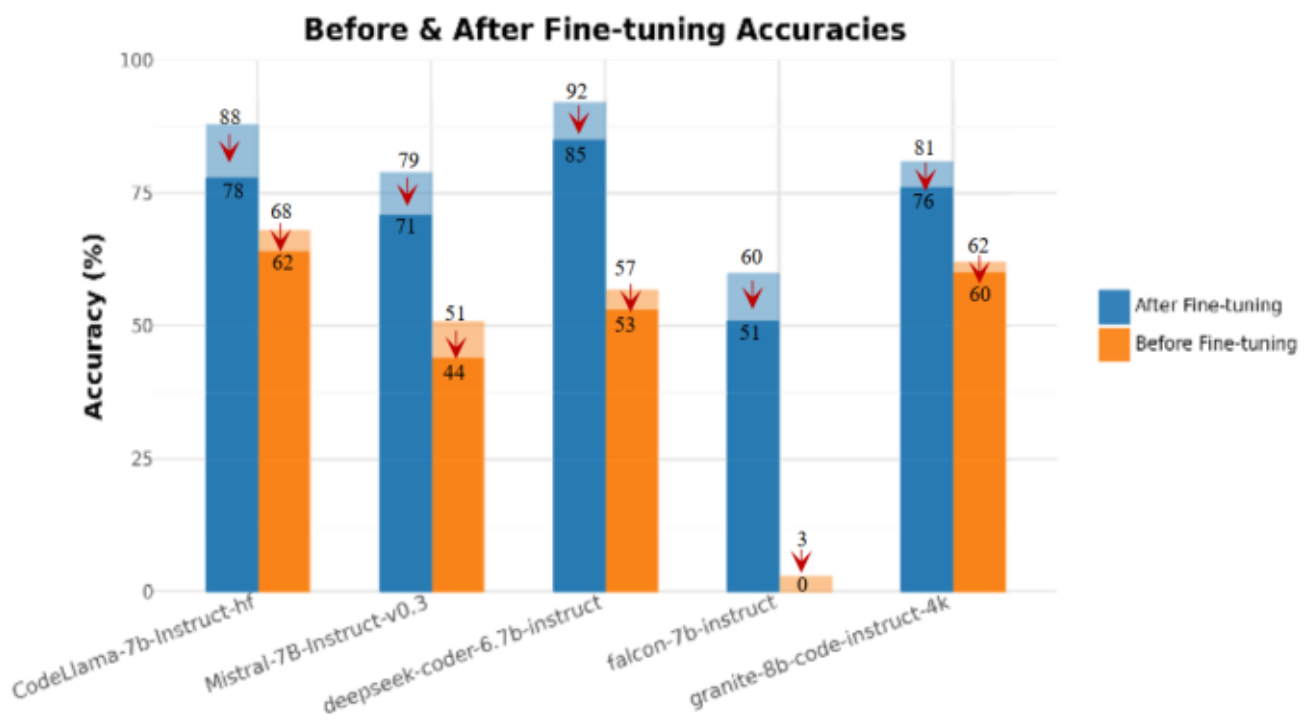

**Fig. 9.** Generall accuarcy in generating synthesizable circuits. Lighter colors represent syntactic correctness

Following this, the fine-tuned LLMs were evaluated on the 15¬eval set. This step offered a more detailed insight into the performance of each LLM and revealed potential directions for future improvements.

As shown in Table 3, most LLMs were able to generate basic functional VHDL modules successfully after fine-tuning. However, as design complexity increased, the occurrence of errors also increased noticeably.

For basic circuits including the 9-to-1 multiplexer, D Flip Flop, Priority Encoder, Half-adder, and XNOR gate, success rates were generally high across all models. These fundamental digital components represent the building blocks that appeared frequently in the training dataset, allowing models to learn their patterns effectively. The 9-to-1 multiplexer, for instance, achieved 100% success rate on syntax checking across all models except Falcon.

Intermediate complexity circuits presented more challenges. The PWM generator, RAM module, and FSM designs showed increased failure rates, particularly in functional verification. The PWM and 8-bit comparator implementation proved especially challenging, with functional success rates dropping to around 40-60% for most models.

Advanced circuits pushed all models to their limits. The 8-bit FFT design task saw complete failure in generating syntactically correct VHDL implementation across all models. The matrix multiplier circuit also proved difficult, that only DeepSeek managed to generate working designs but with low accuracy.

Finally, as shown in Table 4, hardware prototyping as the 3rd level of checking. It can be seen even though all models have been fine-tuned using identical dataset, their outputs for same design task would vary where some LLMs would generate more optimized VHDL code. The data presented in this table provides additional insights into the level of design optimality that can be expected from these LLMs when fine-tuned with the curated dataset. Further steps taken during hardware prototyping was configuring constraints files and mapping IOs of circuit to physical switches and LEDs on Zedboard. This was done on certain design tasks like PWM generation, counters and similar to confirm that design works on real hardware.

As previously discussed, the primary limitation of both the proposed dataset and the fine-tuned models lies in their ability to effectively generate complete designs confined to a single *.vhd* file. In practical scenarios, more complex digital systems are typically structured across multiple VHDL files, where individual components and entities are defined separately and interconnected through a dedicated top-level module.

This outcome reflects the composition of the fine-tuning dataset, which was primarily made up of simpler examples and lacked uniformity in design complexity. This imbalance was further evidenced by the observation that outputs for simpler tasks were often identical, suggesting that the models had potentially overfit to these overrepresented examples. In contrast, outputs for more complex tasks showed much more variation in VHDL code itself, indicating insufficient training examples for these patterns. This biased distribution negatively impacted the model's ability to generalize well when generating sophisticated designs, as the fine-tuning process was inherently skewed toward simpler examples that dominated the dataset.

| Design task | | CodeLlama | | | Mistral | | | Granite-code | | | Falcon | | | DeepSeek-coder | | |
|---|---|---|---|---|---|---|---|---|---|---|---|---|---|---|---|---|
| | | Syntax GHDL | Synthesis GHDL | Func-TB | Syntax GHDL | Synthesis GHDL | Func-TB | Syntax GHDL | Synthesis GHDL | Func-TB | Syntax GHDL | Synthesis GHDL | Func-TB | Syntax GHDL | Synthesis GHDL | Func-TB |
| **Basic** | 9to1 multiplexer | 10/10 | 10/10 | 10/10 | 10/10 | 10/10 | 10/10 | 10/10 | 10/10 | 10/10 | 10/10 | x | x | 10/10 | 10/10 | 10/10 |
| | D Flip Flop | 10/10 | 10/10 | 10/10 | 10/10 | 10/10 | 10/10 | 10/10 | 10/10 | 10/10 | 10/10 | 10/10 | 10/10 | 10/10 | 10/10 | 10/10 |
| | Priority Encoder LSB | 10/10 | 10/10 | 10/10 | 10/10 | 10/10 | 10/10 | 10/10 | 10/10 | 10/10 | 10/10 | 10/10 | 10/10 | 10/10 | 10/10 | 10/10 |
| | Half-adder | 10/10 | 10/10 | 10/10 | 10/10 | 10/10 | 10/10 | 10/10 | 10/10 | 10/10 | 10/10 | 10/10 | 10/10 | 10/10 | 10/10 | 10/10 |
| | XNOR gate | 10/10 | 10/10 | 10/10 | 10/10 | 10/10 | 10/10 | 10/10 | 10/10 | 10/10 | x | x | x | 10/10 | 10/10 | 10/10 |
| **Intermediate** | 4bit_counter | 10/10 | 10/10 | 10/10 | 9/10 | 9/10 | 9/10 | 10/10 | 10/10 | 10/10 | 9/10 | 9/10 | 9/10 | 10/10 | 10/10 | 10/10 |
| | Single port RAM | 0/10 | x | x | 10/10 | x | x | 6/10 | 6/10 | 3/10 | 10/10 | 10/10 | 10/10 | 0/10 | x | x |
| | FSM | 9/10 | 9/10 | 8/10 | 10/10 | 10/10 | 7/10 | 4/10 | 4/10 | 4/10 | 5/10 | 5/10 | 4/10 | 10/10 | 10/10 | 10/10 |
| | 8bit comparator | 1/10 | 1/10 | 1/10 | 0/10 | x | x | 10/10 | 1/10 | 0/10 | 0/10 | x | x | 3/10 | 3/10 | 2/10 |
| | PWM generator | 10/10 | 3/10 | 3/10 | 10/10 | 10/10 | 10/10 | 6/10 | 6/10 | 6/10 | 1/10 | 1/10 | x | 10/10 | 10/10 | 10/10 |
| **Advanced** | Matrix multiplier | 0/10 | x | x | 0/10 | x | x | 0/10 | x | x | 0/10 | x | x | 6/10 | 3/10 | 3/10 |
| | AXI stream FIFO | 2/10 | 2/10 | 2/10 | 1/10 | x | x | 3/10 | 3/30 | x | 2/10 | x | x | 10/10 | 10/10 | 7/10 |
| | ReLU_act_func | 10/10 | 10/10 | 10/10 | 7/10 | 7/10 | 7/10 | 1/10 | 1/10 | 1/10 | 10/10 | 10/10 | 8/10 | 10/10 | 10/10 | 10/10 |
| | 8bit FFT | 0/10 | x | x | 0/10 | x | x | 0/10 | x | x | 0/10 | x | x | 0/10 | x | x |
| | Signed addition overflow | 10/10 | 10/10 | 10/10 | 10/10 | 10/10 | 10/10 | 10/10 | 10/10 | 10/10 | 10/10 | 9/10 | 6/10 | 8/10 | 8/10 | 8/10 |
| **Success rate [%]** | | 68 | 63 | 62 | 71 | 64 | 62 | 66 | 60 | 57 | 58 | 49 | 45 | **78** | **76** | **73** |

TABLE III. SYNTAX, SYNTHESIS AND FUNCTIONALITY ASSESSMENT OF FINE-TUNED LLMS ACROSS 15 DESIGNS

| Design task | | CodeLlama | | | Mistral | | | Granite-code | | | Falcon | | | DeepSeek-coder | | |
|---|---|---|---|---|---|---|---|---|---|---|---|---|---|---|---|---|
| | | Resources | Power (*W*) | Timing (*ns*) | Resources | Power (*W*) | Timing (*ns*) | Resources | Power (*W*) | Timing (*ns*) | Resources | Power (*W*) | Timing (*ns*) | Resources | Power (*W*) | Timing (*ns*) |
| Basic | 9to1 multiplexer | CLB:4 IO:14 | 0.48 | 8.67 | CLB: 4 IO: 14 | 0.48 | 8.67 | CLB: 4 IO:14 | 0.48 | 8.7 | x | x | x | CLB: 4 IO: 14 | 0.48 | 8.7 |
| | D Flip Flop | CLB: 2 IO:4 | 0.26 | 4.79 | CLB: 3 IO: 4 | 0.26 | 4.79 | CLB: 2 IO: 4 | 0.26 | 4.7 | CLB: 2 IO: 4 | 0.26 | 4.79 | CLB: 2 IO: 4 | 0.26 | 4.79 |
| | Priority Encoder LSB | CLB: 1 IO: 6 | 0.52 | 7.1 | CBS: 1 IO: 6 | 0.52 | 7.05 | CLB : 1 IO: 6 | 0.52 | 7.1 | CLB: 1 IO: 6 | 0.52 | 7 | CLB: 1 IO: 6 | 0.52 | 7.1 |
| | Half-adder | CLB: 1 IO: 4 | 0.66 | 6.67 | CLB: 1 IO: 4 | 0.66 | 6.67 | CLB: 1 IO: 4 | 0.66 | 6.67 | CLB: 1 IO: 4 | 0.66 | 6.67 | CLB: 1 IO: 4 | 0.66 | 6.7 |
| | XNOR gate | CLB: 1 IO: 3 | 0.43 | 6.41 | CLB: 1 IO: 3 | 0.43 | 6.41 | CLB : 1 IO: 3 | 0.43 | 6.41 | x | x | x | CLB: 1 IO: 3 | 0.43 | 3.66 |
| Intermediate | 4bit_counter | CLB: 7 IO: 6 | 3.52 | 5.2 | CLB: 7 IO: 7 | 3.68 | 4.93 | CLB: 7 IO: 7 | 3.68 | 4.93 | CLB: 7 IO: 7 | 3.68 | 4.93 | CLB: 15 IO: 7 | 0. 05 | 5.13 |
| | Single-port RAM | x | x | x | CLB: 60 IO:24 | 3.24 | 8.85 | CLB: 2 IO: 25 | 8.9 | 7.7 | CLB: 2 IO: 25 | 8 | 7.7 | x | x | x |
| | FSM | CLB: 6 IO: 5 | 0.21 | 6.7 | CLB: 5 IO:5 | 1.21 | 6.84 | CLB: 6 IO: 5 | 0.16 | 6.7 | CLB: 6 IO: 5 | 0.163 | 5 | CLB: 7 IO: 5 | 0.09 | 7.2 |
| | 8bit comparator | CLB: 6 IO: 16 | 0.15 | 4.86 | x | x | x | x | x | x | x | x | x | CLB: 7 IO: 18 | 0.16 | 5 |
| | PWM generator | x | x | x | CLB: 38 IO: 16 | 2.98 | 7.1 | CLB: 27 IO: 14 | 1.38 | 9 | CLB: 24 IO: 14 | 0.732 | 5 | CLB: 22 IO: 14 | 1.2 | 8.9 |
| Advanced | Matrix multiplier | x | x | x | x | x | x | x | x | x | x | x | x | CLB: 10 IO: 70 | 0.12 | 5.38 |
| | AXI steam FIFO | CLB: 7 IO: 6 | 1.44 | 7.17 | x | x | x | CLB: 8 IO: 6 | 4.38 | 5.9 | x | x | x | CLB: 11 IO: 6 | 2.026 | 7.12 |
| | ReLU_act_func | CLB: 10 IO: 18 | 1.36 | 5.11 | CLB: 9 IO: 18 | 1.36 | 5.11 | CLB: 9 IO: 18 | 2 | 5.2 | CLB: 10 IO: 11 | 1.84 | 4.78 | CLB: 10 IO: 18 | 1.36 | 5.11 |
| | 8bit-FFT | x | x | x | x | x | x | x | x | x | x | x | x | x | x | x |
| | Signed addition overflow | CLB: 9 IO: 25 | 5.81 | 8.92 | CLB: 9 IO: 25 | 5.57 | 8.74 | CLB: 8 IO: 25 | 5.31 | 4.68 | CLB: 17 IO: 25 | 6.06 | 11 | LUT: 9 IO: 25 | 5.6 | 8.74 |

TABLE IV. VIVADO'S DEFAULT RESOURCE, POWER AND TIMING REPORT OF FUNCTIONALLY CORRECT DESIGNS IMPLEMENTED ON ZYNQ SOC

## V. DISCUSSION AND FUTURE WORK

The primary challenges encountered in this work highlight important areas for future research:

**Dataset Construction Challenges:**

- VHDL lacks the large, diverse, publicly available corpus found in mainstream software programming languages
- Advanced hardware designs remain proprietary and protected as intellectual property, limiting training data availability
- This data scarcity presents significant barriers for researchers and students outside the hardware industry
- Synthetic code generation may serve as a viable alternative for dataset expansion

**Evaluation Infrastructure Limitations:**

- Manual testbench creation and simulation execution proved extremely labor-intensive, consuming approximately 90% of evaluation time
- While syntax and synthesis checks can be automated through Python scripts, functional assessment requires substantial manual effort
- Larger-scale studies need enhanced automated frameworks for comprehensive evaluation

**Research Community Needs:**

- Development and open-sourcing of standardized evaluation benchmarks for HDL generation, such as CVDP [35], would significantly benefit the research community

**Scope Limitations and Future Directions:**

This work focused on generating functionally correct VHDL designs rather than optimization or advanced verification aspects. Several key limitations point to important future research directions: (1) **Design optimization** - future work should incorporate area, power, and timing objectives beyond basic resource utilization metrics; (2) **Formal verification** - integration of verification constraints, assertions, and comprehensive test suites into both training data and evaluation pipelines; (3) **Hierarchical design support** - extending beyond single-file modules to multi-component systems with inter-module dependencies; and (4) **Dataset diversity** - addressing model overfitting to simple patterns through active learning and synthetic data generation to better represent complex design patterns.

The single-file limitation observed in our results highlights the need for techniques supporting hierarchical design generation and management as seen in [37]. Additionally, while functional verification used custom testbenches, automated test generation and systematic coverage analysis would provide more comprehensive design assessment for larger-scale studies.

## VI. CONCLUSION

This paper presented a systematic investigation into the use of large language models (LLMs) for automating digital circuit design via hardware description languages. We fine-tuned five open-source 7B-parameter LLMs using a curated instruction-based VHDL dataset, employing the QLoRA method for efficient adaptation. The results demonstrated that fine-tuning significantly improved generation quality, with DeepSeekCoder achieving the highest performance gains. To evaluate model outputs, a three-level validation pipeline was introduced, covering syntax correctness, functional behavior, and hardware prototyping. This comprehensive evaluation revealed that syntactic validity alone is insufficient; functional and physical deployment assessments are critical for determining the true utility of generated designs. The successful deployment of LLM-generated circuits on FPGA hardware further confirmed the practicality of this approach. These findings suggest that LLMs, when properly adapted and evaluated, can contribute meaningfully to the hardware design automation pipeline, offering a promising direction for future research in EDA and AI co-design.

**Acknowledgment:**

This paper is an Extended version of a paper accepted at FLLM 2026 [38]. This work was conceived, designed, and written entirely by the authors. AI-based language assistants were used only for English language polishing, with all outputs manually reviewed and revised. The authors bear full responsibility for the content's accuracy, originality, and integrity. Furthermore, this work was supported by the European Union's Horizon Europe programme under the FRAME project (Grant Agreement No. 101298951). The authors thank the FRAME consortium partners for their valuable collaboration